# Synthetic high-order PT symmetry in a single coil resonator

Chao Zeng[1], Zhiwei Guo[1,*], Yong Sun[1,†], Guo Li[1], Kejia Zhu[2], Jun Jiang[3], Yunhui Li[1], Haitao Jiang[1], Yaping Yang[1], Hong Chen[1,††]

[1]MOE Key Laboratory of Advanced Micro-structured Materials, School of Physics Sciences and Engineering, Tongji University, Shanghai 200092, China
[2]Department of Electrical Engineering, Tongji University, Shanghai 201804, China
[3]School of Automotive Studies, Tongji University, Shanghai 210804, China

*2014guozhiwei@tongji.edu.cn
†yongsun@tongji.edu.cn
††hongchen@tongji.edu.cn

Abstract

The exploration of non-Hermitian systems with parity-time (*PT*) symmetry has witnessed immense research interest both fundamentally and technologically in a wide range of subject areas in physics and engineering. One significant example of the principal emerging fields in this context is the *PT* symmetric wireless applications using multiple coils that are spatially separated but mutually coupled with position-dependent coupling strength. Such a spatial *PT* configuration limits the flexibility and miniaturization of the *PT* symmetric designs. As far as this is concerned, inspired by scattering induced two opposite whispering-gallery (WG) modes in an optical resonator, analogously here we experimentally demonstrate a specially constructed second-order (2-nd order) *PT* symmetry in a single coil resonator, whose currents with two different directions are excited by internal bypass capacitor. Our proposed structure has the following peculiar feature: First, the bypass capacitor induces coupling in spectral resonances allow us to observe a 2-nd order phase transition between symmetry regimes, without the need of a second coil in the spatial *PT* case. Under this circumstance, this specially constructed *PT* symmetry can be regarded as synthetic *PT* symmetry, which is enabled by coupling modes with different directions. Second, by introducing two or more internal bypass capacitors, the synthetic high-order *PT* symmetric system bearing such as third-order exceptional point (EP3) in a single coil resonator can be realized. These results will provide a new paradigm to realize higher-order *PT* symmetry towards the investigation of non-Hermitian physics in a synthetic perspective, which can be extended to other physical platforms such as optics and acoustics.

Keywords: Synthetic high-order *PT* symmetry, high-order exceptional point

## I. INTRODUCTION

A growing interest in investigating the quantum mechanical concept of parity-time (*PT*) symmetry has emerged, showing that non-Hermitian systems can exhibit entirely real spectrum before a certain phase-transition point [1]. When the spontaneous breaking of *PT* symmetry happens, namely undergoing a phase transition near an exceptional point (EP) [2,3], the real spectrum turns into a complex one. Recently, *PT* symmetric concept has triggered immense research interest in optics and photonics to investigate certainly important aspects of non-Hermitian systems, as it can utilize the interplay between gain/loss and the coupling strength to manipulate light waves with desired properties [4-11]. These novel discoveries include enhanced sensing [12-21], coherent perfect absorption (CPA) [22-25], advanced lasers [26-32] and topological states [33-36] to mention just a few examples. Inspired by optical and photonic schemes, *PT* symmetry in acoustics [37-41], optomechanics [42-44], optoelectronics [45, 46] and electronics [47-60] have been reported very recently with desired purposes, such as sensitive sensors [48-54] and robust wireless power transfer (WPT) [55-60].

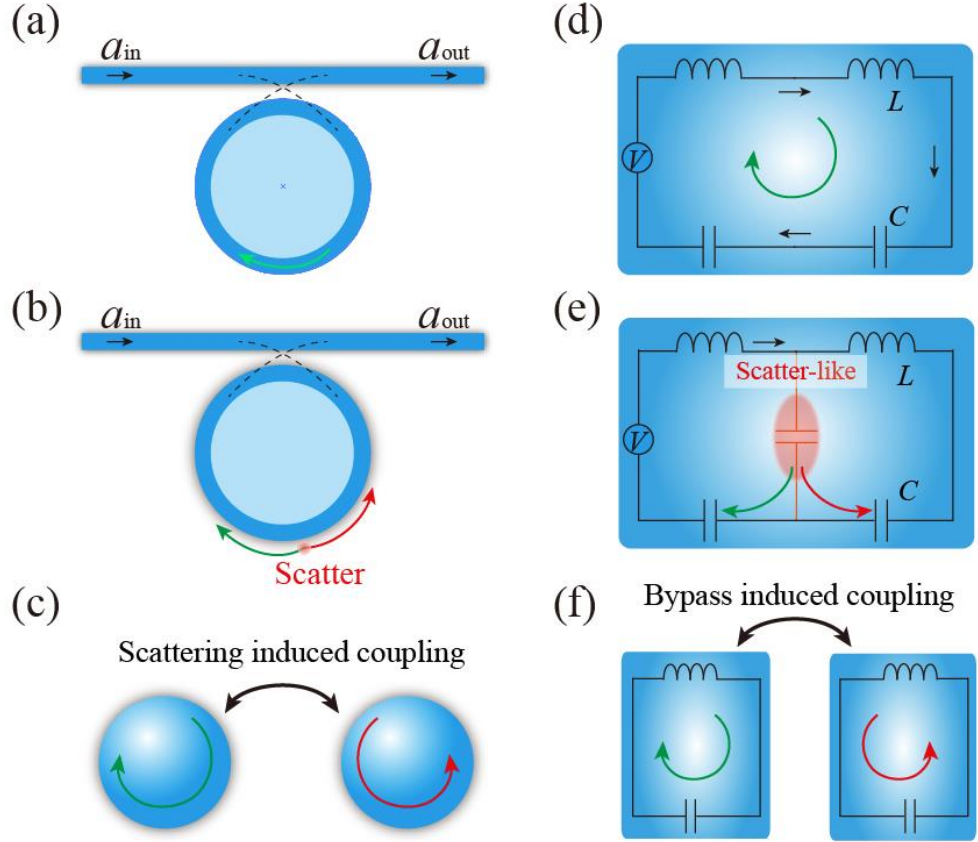

FIG. 1 (a)-(c) Schematic diagrams of the WG structure. (a) A resonator coupled to a waveguide. (b) The directional scattering will take place when a nanoscatterer (or any other form of reflective objects) is placed in the vicinity of the WG structure. (c) Scattering coupling between the CW and CCW waves. (d)-(f) Similar to (a)-(c), but for the synthetic *PT* symmetry in a single coil resonator. The coupling between two degenerated modes are coupling based on the bypass capacitor.

However, realizing the current *PT* symmetry often requires multiple resonators. Especially for higher-order *PT*-symmetric systems, the associated volume problem will limit some practical applications, thus it is desired to construct this *PT* symmetric system using a very simple structure . Whether a new dimension can be introduced into a simple resonance system, solving the above problems is a research topic that people will pay close attention to. As a special kind of resonators, the whispering-gallery (WG) structure has two degenerate modes with different chirality (i.e., the clockwise and the counter-clockwise travelling modes), and its pseudospin has recently been proved to be able to construct *PT* symmetry [26,61-66] or anti-*PT*-symmetry [67]. According to the condition of momentum conservation, one of the chiral modes in WG structure can be selectively excited dependent on the direction of the incident wave [66]. Typically, the structure is depicted schematically in Fig. 1(a). It consists of a single resonator coupled to a waveguide. Within the context of coupled mode theory (CMT), the above structure with the resonant frequency $\omega_0$ only has the clockwise (CW) mode. However, in the presence of scatterer located in the vicinity of the ring resonator shown in Fig. 1(b), the interaction between the scatterer and the evanescent field of the resonator will induce CW and CCW modes. Under this circumstance, it can be considered that two modes are bidirectionally coupled with each other induced by scattering on the resonator surfaces, as shown in Fig. 1(c). It is very interesting that they will scatter the specific chiral WG mode and induce another opposite chiral resonance mode when the particle scatterers are introduced into the WG structure, thus a two-order *PT* symmetry system can be established. Notably, this specially constructed *PT* symmetry can be regarded as synthetic *PT* symmetry, which is enabled by coupling modes with different directions, achieved by imposing scattering in resonant structures such as a whispering-gallery-mode (WGM) cavity.

Inspired by this principle, we theoretically and experimentally demonstrate synthetic high-order *PT* symmetry in a single coil resonator by introducing bypass capacitors. Analogously, the mechanism of internal bypass capacitor induced coupling can also be observed in Figs. 1(d)-1(f). When a bypass capacitor is located in the interior of a *LC* resonator, the currents with two opposite directions are excited. Experimentally, the incident waves excite the left inductance (*L* in the left in Fig. 1(e)) as an effective gain, which is fabricated by lumped circuit elements. But the right inductances (*L* in the right in Fig. 1(e)) are fabricated in a distributed configuration in order to wireless application. When the coupling strength is flexibly controlled by the bypass capacitor, we can clearly observe a second-order (2nd-order) *PT* phase transition between two symmetry regimes, without the need of a second coil in the spatial *PT* case. Likewise, by introducing two or more internal bypass capacitors, we can also construct synthetic high-order (higher than 2nd-order) *PT* symmetric system bearing such as third-order EP (EP3) in a single coil resonator. Unlike the WGM resonator, only two WG modes could be induced by scattering, thus high-order *PT* symmetry is avoided. These results will provide a new paradigm to realize higher-order *PT* symmetry towards the investigation of non-Hermitian physics in a synthetic perspective in optical, acoustic or other frequency bands.

## II. SYNTHETIC SECOND-ORDER *PT* SYMMETRIC SYSTEM VIA A BYPASS CAPACITOR

We analyze this system in theory considering a real situation shown in Fig. 2(a), where the AC source is close to the left resonator. The system dynamics are given by the coupled mode equations (more details in Supplementary Section A):

$$\begin{pmatrix} \omega_{11} + \kappa_1 + i\gamma_1 - \omega & \kappa_1 \\ \kappa_2 & \omega_{22} + \kappa_2 - i\gamma_2 - \omega \end{pmatrix} \begin{pmatrix} a_1 \\ a_2 \end{pmatrix} = \begin{pmatrix} 0 \\ 0 \end{pmatrix}, \qquad (1)$$

where $a_n$ (*n*=1,2) is the amplitude of the resonator with resonant frequency $\omega_{nn} = 1/\sqrt{L_n C_n}$, respectively. The coupling rate is $\kappa_n = \frac{1}{2\omega_{nn} L_n C_0}$ and the loss rate is $\gamma_n = \frac{R_n}{2L_n}$, respectively. Equation (1) describes the generalized 2nd-order non-Hermitian condition, which can be obtained directly from circuit analysis using the appropriate simplifying approximations. Considering the symmetric condition $L_n = L$ and $R_n = R$, defining $\omega_0 = 1/\sqrt{LC}$, $C_n = \frac{C_0 - C}{CC_0}$ and solving Eq. (1), the eigenfrequencies are

$$\omega = \omega_0 \pm \sqrt{\kappa^2 - \gamma^2}, \qquad (2)$$

where $\gamma = \frac{R}{2L}$ and $\kappa = \frac{\omega_0}{2}\frac{C}{C_0}$. Equation (2) indicates that the ideal 2nd-order $PT$ symmetry has been established. Two eigenfrequencies merge at $\omega_0$ with the critical value on condition that $\gamma = \kappa$ is satisfied. Thus, the second-order EP (EP2) are observed. Besides, if we define $\omega_0 = 1/\sqrt{LC}$, $C_n = C$, solving Eq. (1), the eigenfrequencies are
$$\omega' = \omega_0 + \kappa \pm \sqrt{\kappa^2 - \gamma^2}. \tag{3}$$
Equation (3) indicates non-ideal 2nd-order $PT$ symmetry has been established.

We implement the synthetic 2nd-order $PT$ symmetric system by using the radiofrequency circuit. This system looks like a single coil spatially shown in Figs. 2(b) and 2(c). In the following experiments controlling $L_n = L$, what is unusual is that one lumped inductor with inductance $L_1$ is fabricated using toroidal FeSiAl inductor (S106125, 27mm), the other distributed inductor with inductance $L_2$ is fabricated as a coil using Litz wires by turning $n$=25 with 0.078 mm×400 strands and attaching tightly to the polymethyl methacrylate (PMMA) hollow circular cylinder with outside diameter of $D$=60 cm. Besides, lumped-metallized polyester film capacitors (the withstand voltage more than 1500V) are used as the electronic components with capacitance $C_0$ and $C_1$, which are tuned to resonant frequency $f_0 = \frac{1}{2\pi\sqrt{LC}}$. We measure relevant electrical parameters by using the precision $LCR$ digital bridge (AT2818, Applient) as follows: $L \approx 0.737$ mH, $C \approx 4.76$ nF and $f_0 \approx$85 kHz. $C_0$ is modulated from 10 nF to 140 nF, as shown in Fig. 8. To measure reflection or transmission spectrum, the source and resistance in the circuit diagram are connected to Port 1 and Port 2 of the vector network analyzer in the experiments (Keysight E5071C, source impedance is 50 Ω), respectively. Figure 3 gives reflection coefficient ($S_{11}$) versus frequency for this ideal 2nd-order $PT$ symmetric system. Here, $\gamma$ is fixed to 5.7 kHz and $\kappa$ is varied from 9.1 kHz to 4 kHz, which corresponds to a change in $C_0$ and $C_1$ from 23 nF to 55.5 nF and from 6.06 nF to 5.17 nF, respectively. Figures 4(a) and 3(b) plot the experimental (symbols) and calculated (lines) values of real eigenfrequencies as a function of the parameter $\kappa$ in the proposed circuit. We find that theoretical calculations are in a good agreement with experimental data.

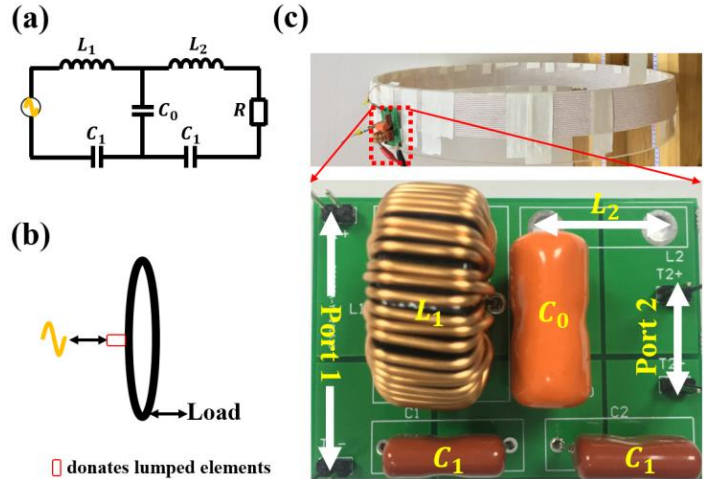

FIG. 2 (a) Circuit diagram of a synthetic second-order (2nd-order) $PT$ symmetric system, where $L_n = L$ and $C_1 = \frac{CC_0}{C_0 - C}$. (b) Corresponding schematic diagram. (c) Corresponding photograph of the experimental sample.

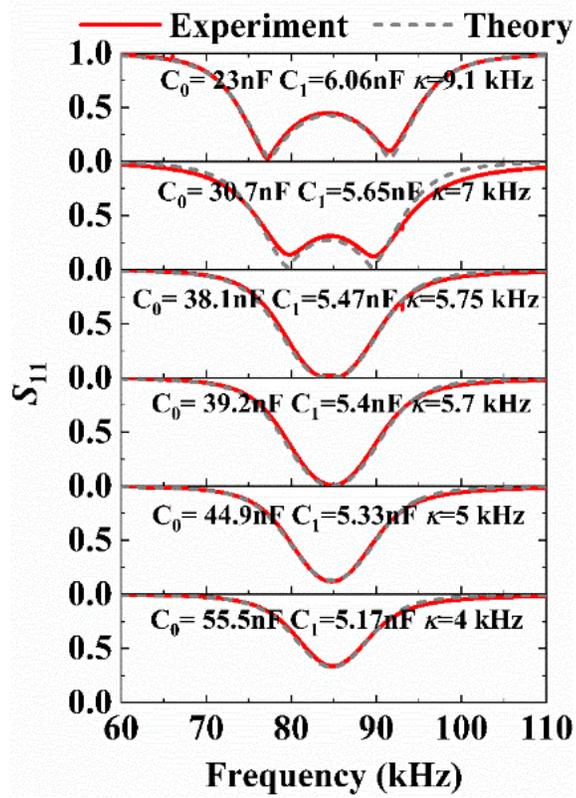

FIG. 3 Evolvement of reflection spectra for this 2-nd $PT$ symmetric system with the coupling strength $\kappa$ decreased from 9.1 kHz to 4 kHz, corresponding $C_0$ and $C_1$ increased from 23 nF to 55.5 nF and reduced from 6.06 nF to 5.17 nF, respectively. Experimental and theoretical results are denoted by solid and dashed lines, respectively.

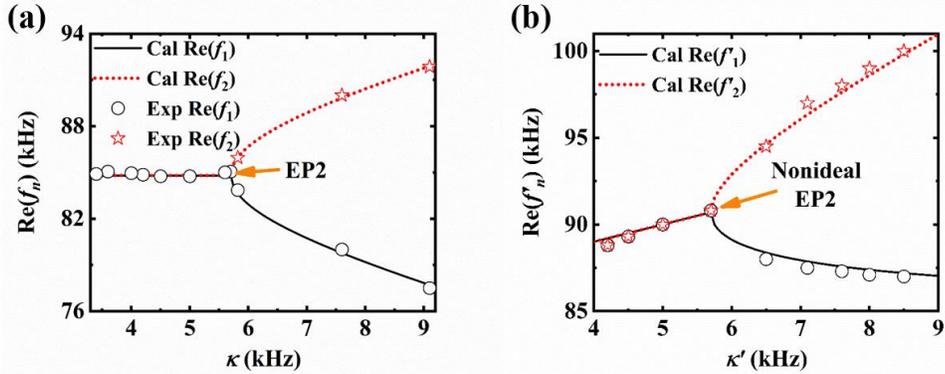

FIG. 4 (a-b) Evolution of the real parts of eigenfrequencies as a function of the equivalent coupling strength $\kappa$ 2nd-order $PT$ symmetric system, for the (a) ideal $PT$ symmetry where $f_0 = \frac{1}{2\pi\sqrt{LC}}$ and (b) non-ideal $PT$ symmetry where $f'_0 = \frac{1}{2\pi\sqrt{LC_n}}$. Experimental and theoretical results are denoted by markers and lines, respectively.

# III. SYNTHETIC HIGH-ORDER $PT$ SYMMETRIC SYSTEM VIA TWO BYPASS CAPACITORS

Generally speaking, experimentally realizing the high-order $PT$ symmetry with EP3 has proven extremely challenging, which needs three identical resonance coils placing coaxially. For example, considering the intrinsic loss of the resonant coil, two adjacent coils need to be very close to each other so that the intrinsic loss can be almost negligible. If do that, the next-nearest-neighbor inductive coupling between transmitter and receiver coils is inevitable.

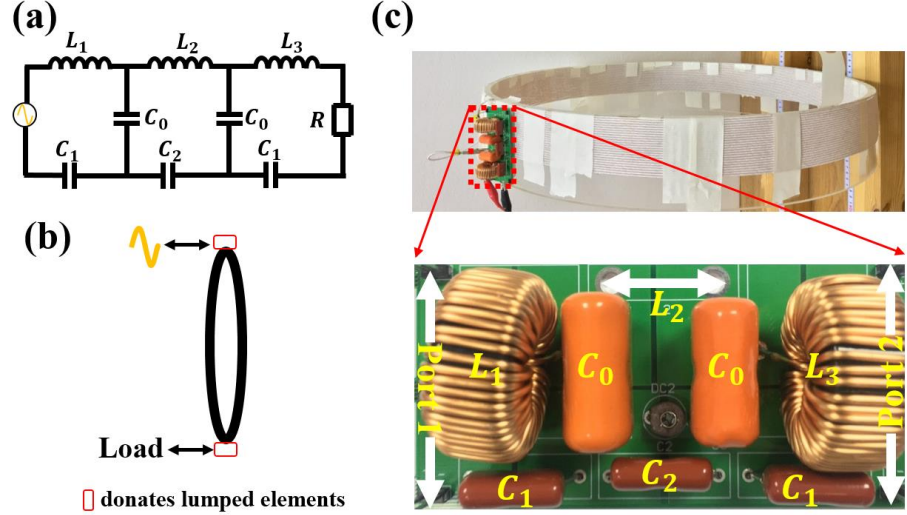

FIG. 5 (a) Circuit diagram of a synthetic third-order (3rd-order) $PT$ symmetric system, where $L_n = L$, $C_1 = \frac{CC_0}{C_0 - C}$ and $C_2 = \frac{CC_0}{C_0 - 2C}$. (b) Corresponding schematic diagram. (c) Corresponding photograph of the experimental sample.

In the following, we will present a new third-order (3rd-order) $PT$ symmetric system based on the coupling of a synthetic cavity with two particles-like and a normal resonator. Here the equivalent circuit model is shown in Fig. 5(a), which also looks like a single coil spatially shown in Figs. 5(b) and 5(c). Similarly, though we control $L_n = L$, two lumped inductors are fabricated with inductance $L_1$ and $L_3$, another distributed inductor is fabricated with inductance $L_2$. Besides, lumped metalized polyester film capacitors are used as the electronic components with capacitance $C_0$ and $C_n$, which are tuned to resonant frequency $f_0 = \frac{1}{2\pi\sqrt{LC}}$. Under these circumstances, the next-nearest-neighbor inductive coupling between $L_1$ and $L_3$ is considered to be ignored. Figure 6 gives reflection coefficient ($S_{11}$) versus frequency for this 3nd-order $PT$ symmetric system. We find a good agreement between theoretical and measurement results. The eigenfrequencies can be obtained as follows (see more details in Supplementary Section B):

$$\omega_1 = \omega_0, \quad \omega_{2,3} = \omega_0 \pm \sqrt{2\kappa^2 - \gamma}, \tag{4}$$

where $\gamma = \frac{R}{2L}$, $\kappa = \frac{\omega_0}{2}\frac{C}{C_0}$. This equation indicates that when $\gamma$ and $\kappa$ reach to a critical value (in this case, $\gamma = \sqrt{2}\kappa$), all three eigenfrequencies coalesce at $\omega_0$ and the system exhibits the EP3 in the eigenspectrum shown in Fig. 7. This method will also be used to realize higher-order (higher than 3rd-order) *PT* symmetry in a single coil resonator in the long term.

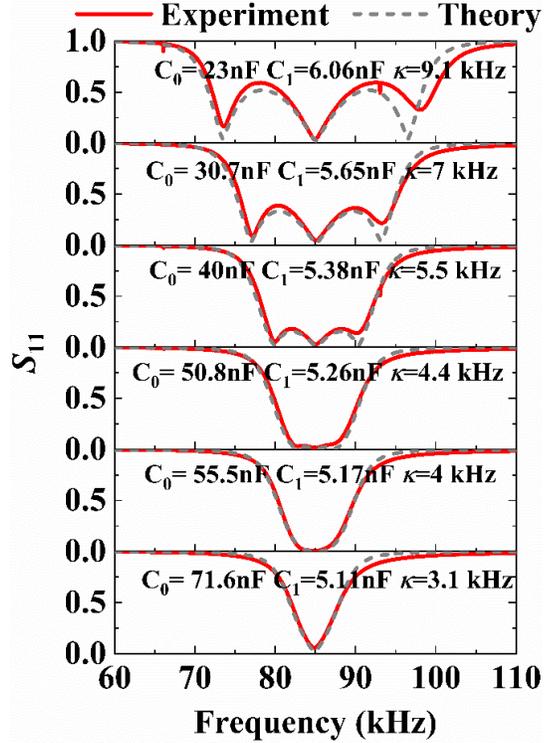

FIG. 6 Evolvement of reflection spectra for this 3$^{rd}$-order *PT* symmetric system with the coupling strength $\kappa$ decreased from 9.1 kHz to 3.1 kHz, corresponding capacitance $C_0$ and $C_1$ increased from 23 nF to 71.6 nF and reduced from 6.06 nF to 5.11 nF, respectively. Experimental and theoretical results are denoted by solid and dashed lines, respectively.

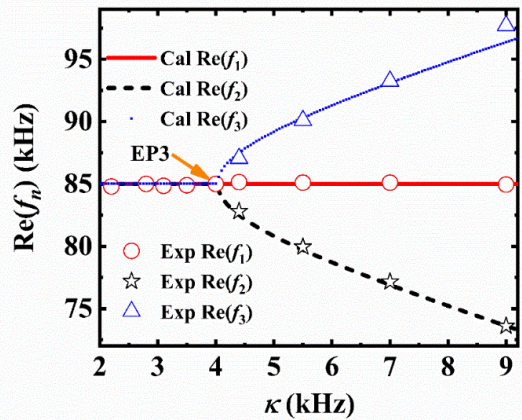

FIG. 7 Evolution of the real parts of eigenfrequencies in the synthetic high-order *PT* symmetric system as a function of the equivalent coupling strength $\kappa$. The solid line describes the calculated result. The symbols give the experimental results.

## IV. CONCLUSION

In conclusion, we have realized a synthetic *PT* symmetry in a single coil resonator spatially. Our work opens a new frontier in the emerging fields in non-Hermitian physics based on higher-order *PT* symmetry with different configurations. Our results also offer a unique perspective in exploring non-Hermitian physics, which can be extended to investigate a wide range of phenomena.

## ACKNOWLEDGEMENTS

This work is sponsored by the National Key Research Program of China (No. 2016YFA0301101), by the Natural Science Foundation of China (Nos. 91850206, 61621001, 11674247, 11974261, 11604136), by the Shanghai Science and Technology Committee (Nos. 18JC1410900 and 18ZR1442900), by the Fundamental Research Funds for the Central Universities, by the Shanghai Super Postdoctoral Incentive Program, by the China Postdoctoral Science Foundation (Grant Nos. 2019TQ0232 and 2019M661605).

## APPENDIX A: METHODS OF SYNTHETIC 2ND-ORDER *PT* SYMMETRY

As is shown in Fig. 2(a), Using Kirchhoff's laws, the voltages and currents are related as

$$\begin{pmatrix} i\omega L_1 + \frac{1}{i\omega C_1} + \frac{1}{i\omega C_0} - R_1 & \frac{1}{i\omega C_0} \\ \frac{1}{i\omega C_0} & i\omega L_2 + \frac{1}{i\omega C_2} + \frac{1}{i\omega C_0} + R_2 \end{pmatrix} \begin{pmatrix} I_1 \\ I_2 \end{pmatrix} = \begin{pmatrix} 0 \\ 0 \end{pmatrix}. \quad (S1)$$

Where $I_n$, $n=1,2$ is the electric current with different direction, $R_1$ and $R_2$ are impedance of source and load, respectively. Considering the electrodynamic potential of AC source $U = -I_1 R_1$, the non-Hermitian condition is achieved when the gain and loss parameters, namely $-R_1$ and $R_2$, are defined. By substituting the amplitude $a_n = -iL_n \frac{dI_n}{d\tau}$, $\tau = \omega t$ into Eq. (S1), we also have

$$\begin{pmatrix} i\omega_{11}^2 + \frac{i}{2\omega_1 L_1 C_0} + \frac{R_1 \omega}{2L_1} - i\omega^2 & \frac{i}{2\omega_1 L_1 C_0} \\ \frac{i}{2\omega_2 L_2 C_0} & i\omega_{11}^2 + \frac{i}{2\omega_1 L_1 C_0} - \frac{R_2 \omega}{2L_1} - i\omega^2 \end{pmatrix} \begin{pmatrix} a_1 \\ a_2 \end{pmatrix} = \begin{pmatrix} 0 \\ 0 \end{pmatrix}. \quad (S2)$$

We make the approximation $\omega^2 - \omega_{nn}^2 \approx 2\omega(\omega - \omega_{nn})$, $\omega_{nn} = \frac{1}{\sqrt{L_n C_n}}$, equation (S2) then reduces to

$$\begin{pmatrix} \omega_{11} + \kappa_1 + i\gamma_1 - \omega & \kappa_1 \\ \kappa_2 & \omega_{22} + \kappa_2 - i\gamma_2 - \omega \end{pmatrix} \begin{pmatrix} a_1 \\ a_2 \end{pmatrix} = \begin{pmatrix} 0 \\ 0 \end{pmatrix}. \quad (S3)$$

Where the coupling rate is $\kappa_n = \frac{1}{2\omega_{nn} L_n C_0}$ and the loss rate is $\gamma_n = \frac{R_n}{2L_n}$, respectively. Solving Eq. (S3), the eigenfrequencies are

$$\omega = \frac{\omega_{11} + \kappa_1 + \omega_{22} + \kappa_2}{2} + i\frac{\gamma_2 - \gamma_1}{2} \pm \sqrt{\kappa_1 \kappa_2 + (\frac{\omega_{22} + \kappa_2 - \omega_{11} - \kappa_1}{2} + i\frac{\gamma_2 + \gamma_1}{2})^2}. \quad (S4)$$

Considering the symmetric condition $L_n = L$ and $R_n = R$, defining $\omega_0 = 1/\sqrt{LC}$, $C_n = \frac{CC_0}{C_0 - C}$, the eigenfrequencies are

$$\omega = \omega_0 \pm \sqrt{\kappa^2 - \gamma^2}. \quad (S5)$$

Where $\gamma = \frac{R}{2L}$ and $\kappa = \frac{\omega_0}{2}\frac{C}{C_0}$. In this system, $\kappa$ is related to $C_0$, as shown in Fig. 8.

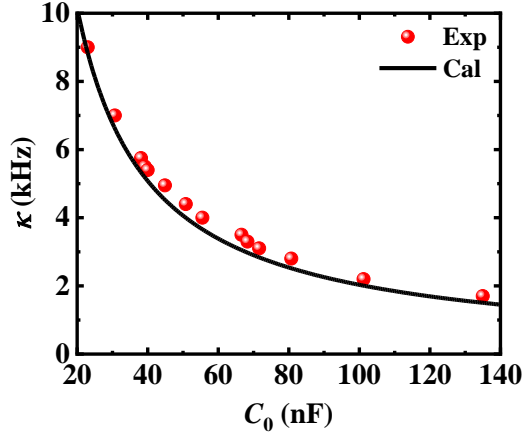

FIG. 8 The parameter $\kappa$ as a function of $C_0$. The solid line describes the calculated result. The red dots signify experimental results.

# APPENDIX B: METHODS OF SYNTHETIC HIGH-ORDER $PT$ SYMMETRY

As is shown in Fig. 3, according to Kirchhoff's current laws, the voltages and currents are given as follows:

$$\begin{pmatrix} i\omega L_1 + \frac{1}{i\omega C_1} + \frac{1}{i\omega C_0} - R_1 & \frac{1}{i\omega C_0} & 0 \\ \frac{1}{i\omega C_0} & i\omega L_2 + \frac{1}{i\omega C_2} + \frac{2}{i\omega C_0} & \frac{1}{i\omega C_0} \\ 0 & \frac{1}{i\omega C_0} & i\omega L_3 + \frac{1}{i\omega C_1} + \frac{1}{i\omega C_0} + R_2 \end{pmatrix} \begin{pmatrix} I_1 \\ I_2 \\ I_3 \end{pmatrix} = \begin{pmatrix} 0 \\ 0 \\ 0 \end{pmatrix}. \quad (S6)$$

To simplify the problem, considering the symmetric condition $L_n = L$ and $R_n = R$, defining $\omega_0 = 1/\sqrt{LC}$, $C_{1,3} = \frac{CC_0}{C_0-C}$ and $C_2 = \frac{CC_0}{C_0-2C}$, similarly, we make the approximation $\omega^2 - \omega_0^2 \approx 2\omega(\omega - \omega_0)$, equation (S9) can be reduced as

$$\begin{pmatrix} \omega_0 + i\gamma - \omega & \kappa & 0 \\ \kappa & \omega_0 - \omega & \kappa \\ 0 & \kappa & \omega_0 - i\gamma - \omega \end{pmatrix} \begin{pmatrix} a_1 \\ a_2 \\ a_3 \end{pmatrix} = \begin{pmatrix} 0 \\ 0 \\ 0 \end{pmatrix}. \quad (S7)$$

Solving Eq. (S8), the eigenfrequencies are

$$\omega_1 = \omega_0, \quad \omega_{2,3} = \omega_0 \pm \sqrt{2\kappa^2 - \gamma}. \quad (S8)$$

Where $\gamma = \frac{R}{2L}$ and $\kappa = \frac{\omega_0}{2}\frac{C}{C_0}$. In this system, all three eigenfrequencies merge at $\omega_0$, providing the critical value condition of $\gamma = \sqrt{2\kappa}$ is satisfied.